\title{Some incidence theorems and integrable discrete equations}
\author{V.E. Adler}
\date{\empty}

\documentclass[a4paper,12pt]{article}
\usepackage{amsmath,amssymb,amsthm,graphicx}
\usepackage[english]{babel}

\def\Integer{\mathbb Z}
\def\Real{\mathbb R}
\def\CP{\mathbb C\mathbb P}
\def\CC{{\cal C}}
\newtheorem{theorem}{Theorem}
\newtheorem{lemma}[theorem]{Lemma}

\begin{document}
\maketitle

\begin{quote}\small
Institut f\"ur Mathematik, Technische Universit\"at Berlin,\\
Str.~des 17.~Juni 136, 10623 Berlin, Germany\\
E-mail: {\tt adler@itp.ac.ru}
\bigskip

{\bf Abstract.} Several incidence theorems of planar projective geometry are
considered. It is demonstrated that generalizations of Pascal theorem due to
M\"obius give rise to double cross-ratio equation and Hietarinta equation.
The construction corresponding to the double cross-ratio equation is a
reduction to a conic section of some planar configuration $(20_3 15_4)$.
This configuration provides a correct definition of the multidimensional
quadrilateral lattices on the plane.
\end{quote}

\let\oldf=\thefootnote \let\thefootnote=\strut
\footnote{On leave from L.D.~Landau Institute for Theoretical Physics,
 Chernogolovka, Russia.}
\footnote{This research was supported by the Alexander von Humboldt
 Foundation.}
\let\thefootnote=\oldf

\section{Introduction}

Accordingly to \cite{ABS,NW}, a $m$-dimensional partial difference equation
$F[x]=0$, $x:\Integer^m\to\Real^d$, is called integrable if it can be
self-consistently imposed on each $m$-dimensional sublattice in
$\Integer^{m+1}$. Assume that this equation can be interpreted as a
geometric construction which defines some elements of a figure by the other
ones. This may be a figure not necessarily in $\Real^d$, for example, $x$
may play the role of parameter on some manifold. Then the integrability
means that some complex figure exists which contains several copies of our
basic figure, and this complex figure can be constructed by the given
elements in several ways. In other words, the self-consistency property is
expressed geometrically as some incidence theorem.

In some examples, the construction of the basic figure is itself possible
due to an incidence, so that already the equation $F[x]=0$ is equivalent to
some incidence theorem. This low-level incidence occurs, when our equation
is a reduction in some more general equation, or, geometrically, our
construction is a particular case of some more general construction.

The present paper illustrates these notions by the examples of double
cross-ratio equation (Nimmo and Schief \cite{NS}), Hietarinta equation
\cite{H}, and quadrilateral lattices (Doliwa and Santini \cite{DS}). Recall
that double cross-ratio, or discrete Schwarz-BKP equation appears in the
theory of B\"acklund transformations for 2+1-dimensional sine-Gordon and
Nizhnik-Veselov-Novikov equations and is equivalent to Hirota-Miwa equation
and star-triangle map. Its applications in geometry were found in
\cite{KS1,KS2}. Quadrilateral lattices can be considered as the discrete
analog of the conjugated nets \cite{DS} and are the object of intensive
study in the modern theory of integrable systems and discrete geometry (see,
for example, the recent review \cite{BMS}).

We start in Section \ref{s:mobius} from the M\"obius theorem on the polygons
inscribed in a conic section and Theorem \ref{th:uv} which is its
modification. Section \ref{s:cr} is devoted to the analytical description of
the figures under consideration. In Sections \ref{s:2cr}, \ref{s:H} we
consider in more details the particular cases of these theorems,
corresponding to double cross-ratio and Hietarinta equations.

In Section \ref{s:pnet} we show that double cross-ratio equation is a
reduction of some more general mapping which corresponds geometrically to
some planar configuration with the symbol $(20_3 15_4)$. In turn, this
configuration is just the projection on the plane of the elementary cell of
the quadrilateral lattice. Integrability of quadrilateral lattices in
$\Real^3$, proved in \cite{DS}, implies the integrability of this mapping on
the plane, and next, reduction to the conic section proves the integrability
of double cross-ratio equation.

In brief, the content of the paper is illustrated by the following diagram.
\medskip

\def\vv{\vrule height1.8em width0em depth1.2em}
\centerline{
\begin{tabular}{ll}
 M\"obius theorem \quad $\stackrel{N=3}{\longrightarrow}$ \quad
 Pascal theorem   \quad $\stackrel{N=3}{\longleftarrow}$  & Theorem \ref{th:uv} \\
  \qquad  $\big\downarrow$ \scriptsize $N=4$ &
  \qquad  $\big\downarrow$ \scriptsize $N=4$ \vv\\
 double cross-ratio eq. (\ref{2cr}) & Hietarinta eq. (\ref{H}) \\
  \qquad $\big\uparrow$ reduction to the conic section & \vv\\
 Theorem \ref{th:60}, quadrilateral lattices on the plane & \\
  \qquad $\big\uparrow$ projection & \vv\\
 quadrilateral lattices in $\Real^d$ &
\end{tabular}}

\medskip

\section{Generalizations of Pascal theorem}\label{s:mobius}

\begin{figure}[htb]
\center{\includegraphics[width=65mm]{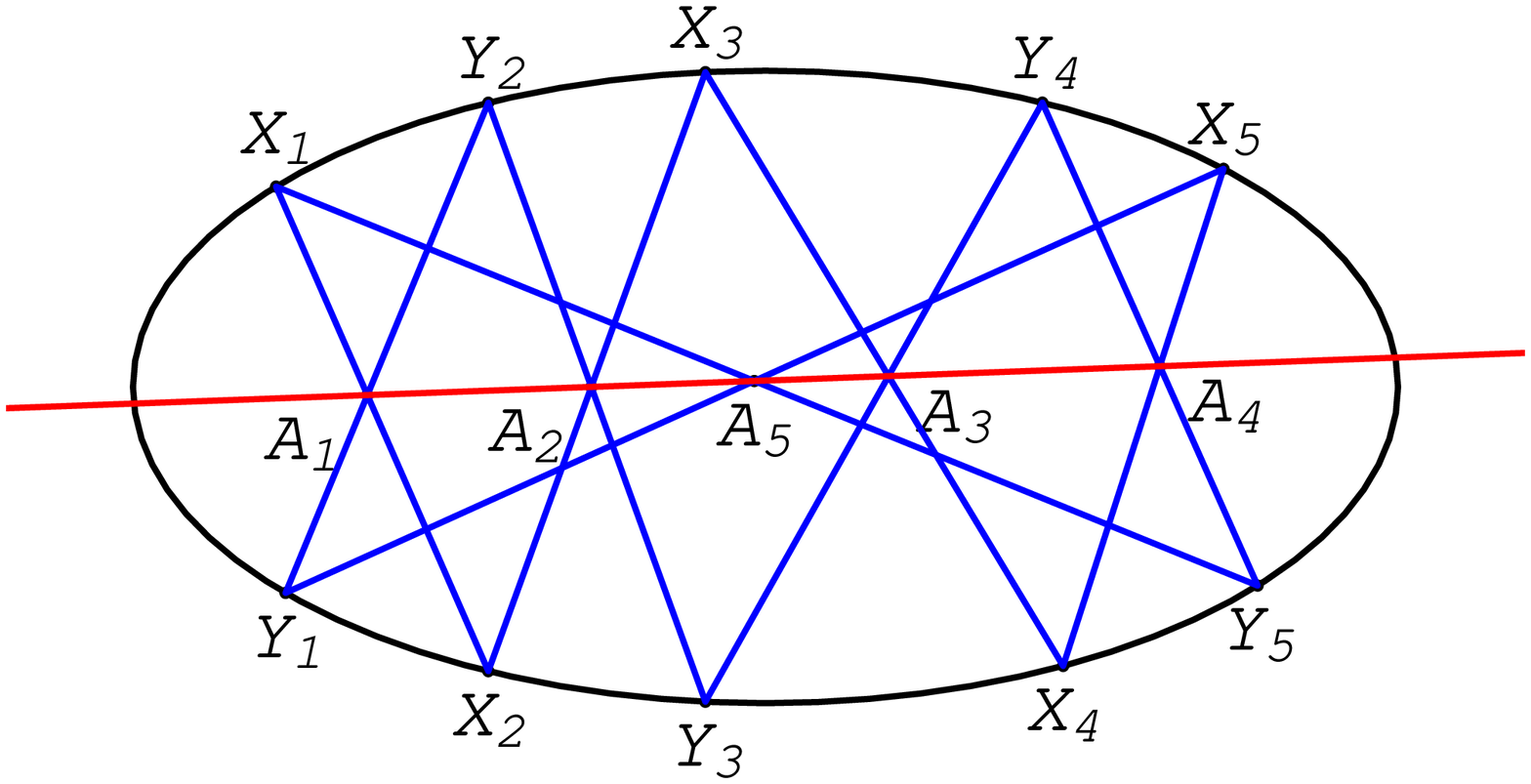}
 \includegraphics[width=65mm]{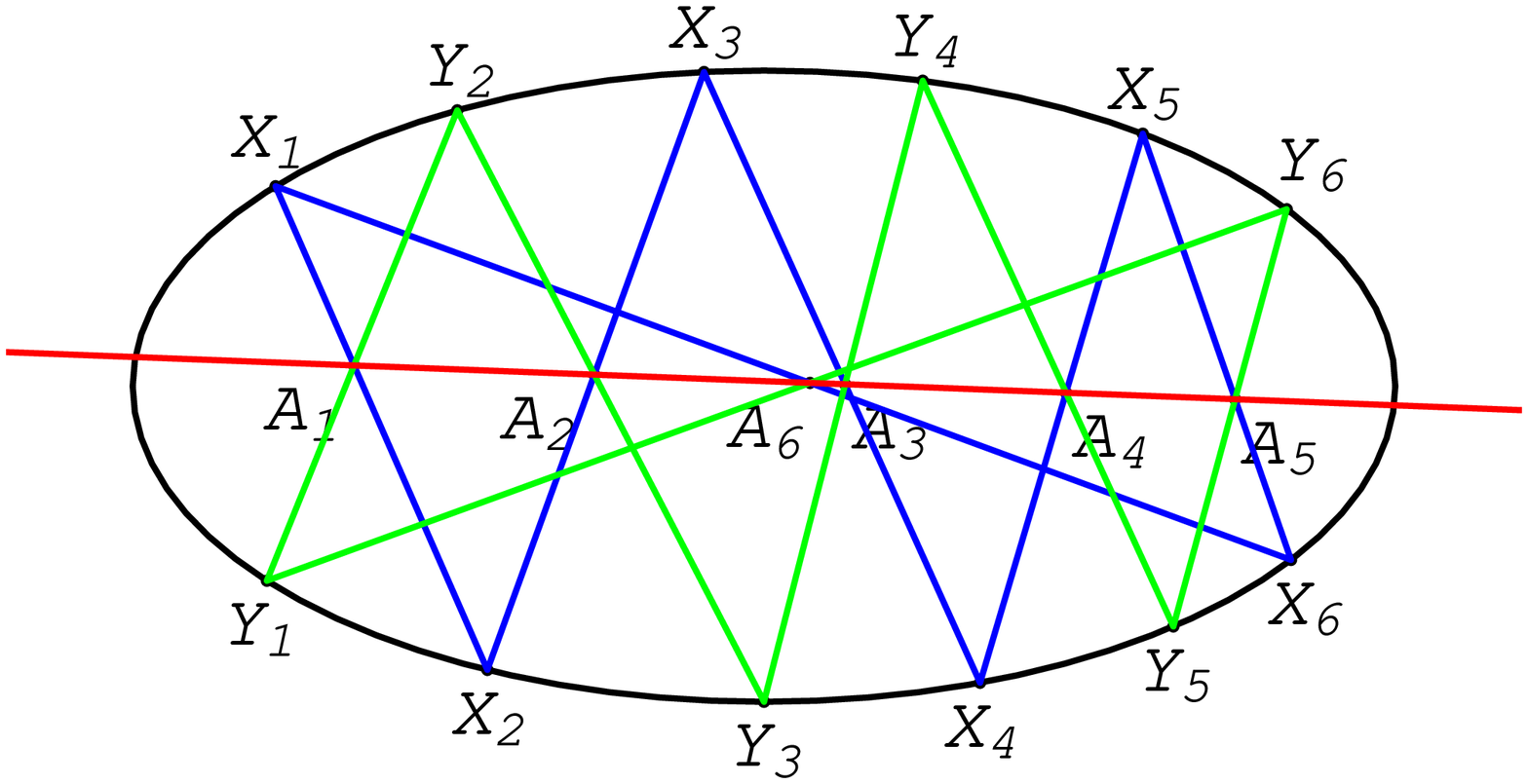}}
\caption{M\"obius theorem, $N=5,6$}\label{fig:x56}
\end{figure}

Recall that, accordingly to Pascal theorem (the prolongations of) the
opposite sides of a hexagon inscribed in a conic section meet on a straight
line. In 1847, M\"obius found the following generalizations of this theorem
(see fig.~\ref{fig:x56}):

1) let $(4n+2)$-gon be inscribed in a conic section and $2n$ pairs of its
opposite sides meet on a straight line, then the same is true for the
remaining pair;

2) let two $2n$-gons be inscribed in a conic section and $2n-1$ pairs of
their corresponding sides meet on a straight line, then the same is true for
the remaining pair.

In order to fix the notations, we reformulate this as follows.

\begin{theorem}[M\"obius]\label{th:mobius}
Let $X_1,Y_1,\dots,X_N,Y_N$ be points on a conic section. Consider the
intersection points $A_j=X_jX_{j+1}\cap Y_jY_{j+1}$, $j=1,\dots N-1$ and
\[
   A_N=\left\{\begin{array}{lll}
   X_NY_1\cap Y_NX_1 & \mbox{\rm if} & N=2n+1,\\
   X_NX_1\cap Y_NY_1 & \mbox{\rm if} & N=2n.
 \end{array}\right.
\]
If all of these points except possibly one are collinear then the same is
true for the remaining point.
\end{theorem}

The proof by M\"obius (based on the Gergonne proof of Pascal theorem) is
very simple. Consider the projective transformation of the plane which maps
the conic section into a circle and sends the line of intersections to
infinity. Then the statement is that if all pairs of the opposite (resp.
corresponding) sides except for possibly one are parallel then this is true
for the remaining pair as well. This easily follows from the fact that a
pair of parallel chords cuts off equal arcs of the circle with opposite
orientation and vice versa: $XX'\;\|\;YY'\ \Leftrightarrow\
\widehat{XY}=-\widehat{X'Y'}$. The change of orientation explains the
difference between the cases of odd and even $N$. (Possibly, this was the
first step on the way to the invention of the M\"obius band, in 1861?)

Another proof can be obtained by applying Pascal theorem to some sequence of
hexagons. Assume that we have to prove the collinearity of the last point
$A_N$. Consider the hexagons $X_1X_2X_3Y_1Y_2Y_3$, $X_1Y_3Y_4Y_1X_3X_4$,
$X_1X_4X_5Y_1Y_4Y_5$ and so on. On each step we prove that some new
intersection point is collinear: first, the point $A'=X_1Y_3\cap Y_1X_3$,
then $A''=X_1X_4\cap Y_1Y_4$, $A'''=X_1Y_5\cap Y_1X_5$ and so on, until we
come to the point $A_N$.

\begin{figure}[htb]
\center{\includegraphics[width=65mm]{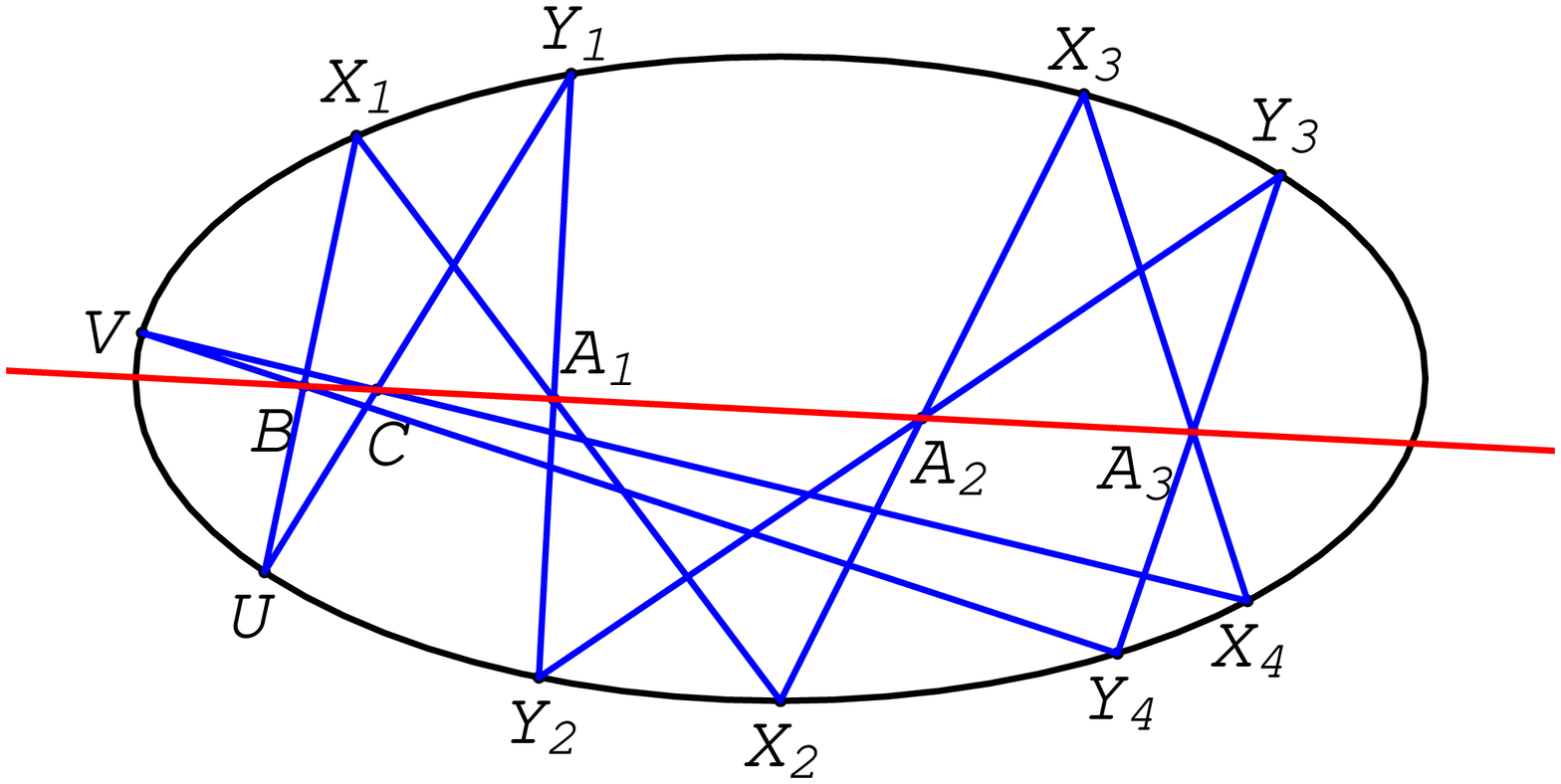}
 \includegraphics[width=65mm]{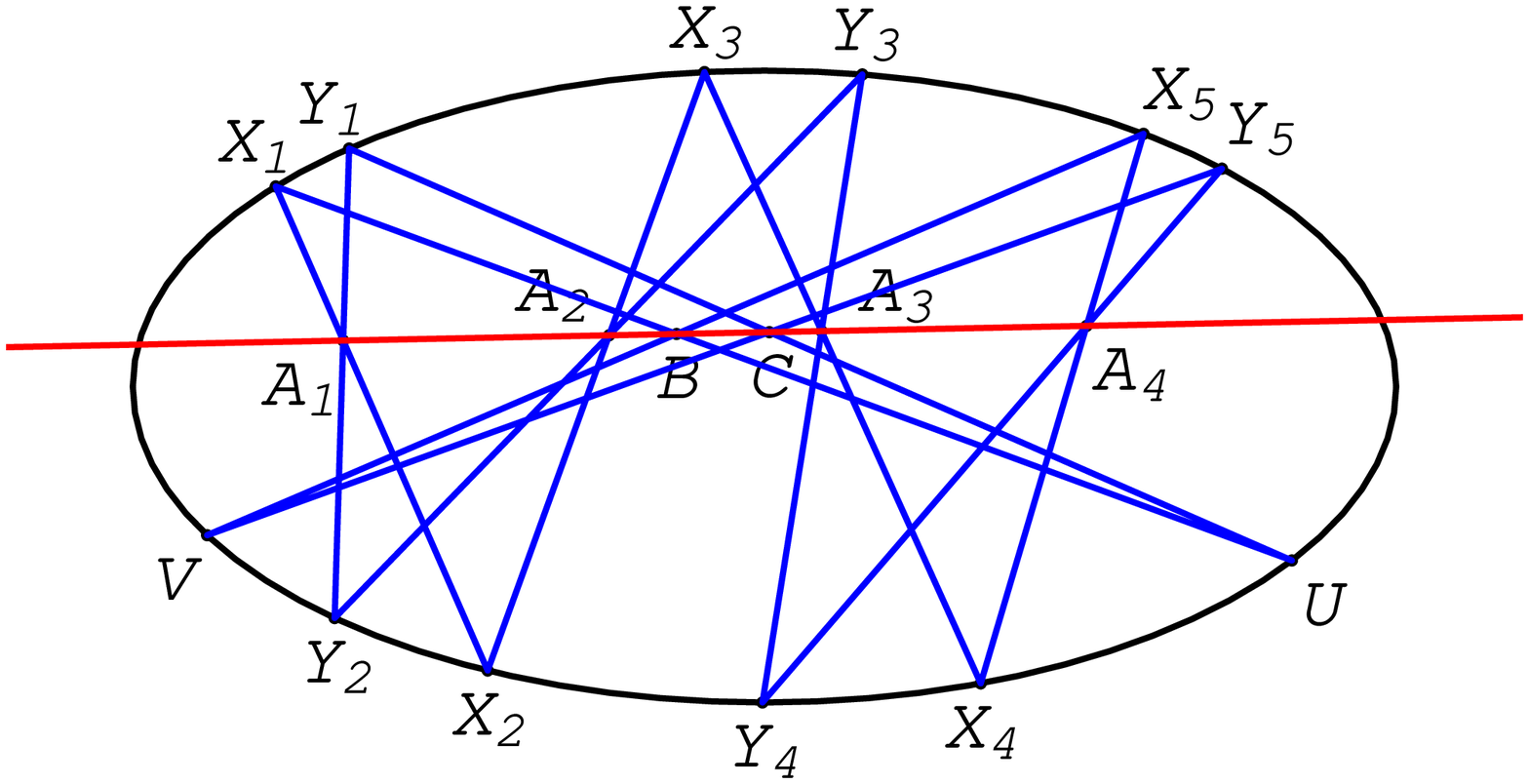}}
\caption{Theorem \ref{th:uv}, $N=4,5$}\label{fig:uv}
\end{figure}

M\"obius theorem admits several variations which can be proved by the same
reasoning. One of them is given by the following statement (see
fig.~\ref{fig:uv}).

\begin{theorem}\label{th:uv}
Let polygon $UX_1\dots X_NVY_N\dots Y_1$ be inscribed in a conic section.
Consider the intersection points $A_j=X_jX_{j+1}\cap Y_jY_{j+1}$, $j=1,\dots
N-1$ and
\begin{align*}
 B &= UX_1\cap Y_{2n}V,  & C &= UY_1\cap X_{2n}V  & \mbox{\rm if }& N=2n, \\
 B &= UX_1\cap X_{2n+1}V,& C &= UY_1\cap Y_{2n+1}V& \mbox{\rm if }& N=2n+1.
\end{align*}
If all of these points except possibly one are collinear then the same is
true for the remaining point.
\end{theorem}

\section{Cross-ratio lattice}\label{s:cr}

Here we consider the system of algebraic equations which is equivalent to
the collinearity of the intersection points. In this language, Theorems
\ref{th:mobius}, \ref{th:uv} mean that any equation of this system is a
consequence of all the others. The derivation of the system is based on the
Lemma \ref{l:48} below. The notation $(a,b,c,d)=(a-b)/(b-c)\cdot(c-d)/(d-a)$
for the cross-ratio is used and, more general, the multi-ratio is denoted as
\[
  (a,b,\dots,c,d)=(a-b)/(b-\dots)\cdots (c-d)/(d-a).
\]

\begin{lemma}\label{l:48}
1) Let the broken lines $X_1X_2X_3X_4$ and $Y_1Y_2Y_3Y_4$ be inscribed into
a conic section $\CC$, and $x_j,y_j$ be the corresponding values of a
rational parameter on $\CC$. Then the collinearity of the intersection
points $A_j=X_jX_{j+1}\cap Y_jY_{j+1}$, $j=1,2,3$ is equivalent to equation
\begin{equation}\label{xycr}
  (x_1,y_2,x_3,y_4)=(y_1,x_2,y_3,x_4).
\end{equation}
2) Let the broken lines $Y_2Y_1UX_1X_2$ and $PVQ$ be inscribed into a conic
section $\CC$, and $x_j,y_j$, $u,v,p,q$ be the corresponding values of a
rational parameter on $\CC$. Then the collinearity of the intersection
points
\[
 A=X_1X_2\cap Y_1Y_2,\quad B=UX_1\cap PV,\quad C=UY_1\cap QV
\]
is equivalent to equation
\[
  (p,u,q,x_2,y_1,v,x_1,y_2)=1.
\]
\end{lemma}
\begin{proof}
Since all nondegenerate conic sections are equivalent modulo projective
transformations of the plane, and all birational parametrizations of the
conic section are M\"obius-equivalent, it is sufficient to check the
formulae for some parametrization of some conic section, for example for the
parabola $X=(x:x^2:1)$. This is a straightforward computation.
\end{proof}

It is easy to see that M\"obius theorem corresponds to the system
\begin{equation}\label{jcr}
  (x_j,y_{j+1},x_{j+2},y_{j+3})=(y_j,x_{j+1},y_{j+2},x_{j+3}),\quad
  j=1,\dots,N-2
\end{equation}
with the periodic boundary conditions
\begin{align*}
 &x_{N+1}=x_1,\quad y_{N+1}=y_1,\quad N=2n, \\
 &x_{N+1}=y_1,\quad y_{N+1}=x_1,\quad N=2n+1.
\end{align*}

In particular, for $N=3$ the system turns into equation
$(x_1,y_2,x_3,x_1)=(y_1,x_2,y_3,y_1)$ which is the identity $\infty=\infty$
(Pascal theorem).

For $N=4$ the system consists from equations (\ref{xycr}) and
$(x_2,y_3,x_4,y_1)=(y_2,x_3,y_4,x_1)$ which are obviously equivalent.

In general case, we obtain one more proof of the M\"obius theorem by
checking that the last equation of the system follows from the others. To
this end, notice that the arguments of cross-ratios can be interchanged in
arbitrary order, simultaneously in left and right hand sides, and
$(a,b,c,d)/(a,b,c,e)=(a,e,c,d)$.

This allows to eliminate $x_2,y_2$ from the first and second equations of
the system (\ref{jcr}), resulting in $(x_1,x_3,y_4,x_5)=(y_1,y_3,x_4,y_5)$.
Geometrically, this means that the point $A'=X_1Y_3\cap Y_1X_3$ is collinear
to $A_3,A_4$, so that we follow the proof by recursive applying of Pascal
theorem as described in the previous section. On the next step, eliminating
$x_3,y_3$ from this new equation and third equation of the system results in
equation $(x_1,y_4,x_5,y_6)=(y_1,x_4,y_5,x_6)$ which is equivalent to the
collinearity of $A'',A_4,A_5$. Repeat this procedure until come to the
equation $(x_1,x_{N-2},y_{N-1},x_N)=(y_1,y_{N-2},x_{N-1},y_N)$ if $N$ is odd
or $(x_1,y_{N-2},x_{N-1},y_N)=(y_1,x_{N-2},y_{N-1},x_N)$ if $N$ is even, as
required.

Analogously, Theorem \ref{th:uv} corresponds to the system
\begin{align}\label{jcr'}
  & (x_j,y_{j+1},x_{j+2},y_{j+3})=(y_j,x_{j+1},y_{j+2},x_{j+3}),
  \quad j=1,\dots,N-3,\\
\label{even}
  & \left\{\begin{array}{ll}
     (y_N,u,x_N,x_2,y_1,v,x_1,y_2)&=1 \\
     (y_1,v,x_1,x_{N-1},y_N,u,x_N,y_{N-1})&=1
    \end{array}\right. \qquad \mbox{\rm if } N=2n, \\
\label{odd}
  & \left\{\begin{array}{ll}
     (x_N,u,y_N,x_2,y_1,v,x_1,y_2)&=1 \\
     (y_1,v,x_1,y_{N-1},x_N,u,y_N,x_{N-1})&=1
    \end{array}\right. \qquad \mbox{\rm if } N=2n+1
\end{align}
where equations (\ref{even}) or (\ref{odd}) are equivalent to the
collinearity of the points $B,C,A_1$ and $B,C,A_N$. The ratio of equations
(\ref{even}) is equivalent to equation
$(x_1,y_2,x_{N-1},y_N)=(y_1,x_2,y_{N-1},x_N)$ while the ratio of equations
(\ref{odd}) is equivalent to $(x_1,y_2,y_{N-1},x_N)=(y_1,x_2,x_{N-1},y_N)$.
In both cases, this can be proved to be a consequence of (\ref{jcr'}).

\section{Double cross-ratio equation}\label{s:2cr}

\begin{figure}[htb]
 \center{\includegraphics[width=65mm]{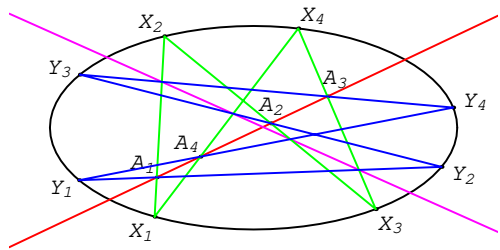}}
 \caption{Intersections in a pair of quadrilaterals}\label{fig:2cr}
\end{figure}

Consider in more details M\"obius theorem at $N=4$. It says that if three
pairs of the corresponding sides of inscribed quadrilaterals $X_1X_2X_3X_4$
and $Y_1Y_2Y_3Y_4$ meet on a straight line, then the same is true for the
fourth pair. As we have seen in the previous section, the figure under
consideration is governed by equation (\ref{xycr})
\[
  (x_1,y_2,x_3,y_4)=(y_1,x_2,y_3,x_4).
\]
Due to the transformation properties of cross-ratio this equation can be
rewritten in several equivalent forms, so that it provides the collinearity
also of several other quadruples of the intersection points (see
fig.~\ref{fig:2cr} where one of these additional lines is shown).

In particular, equation (\ref{xycr}) is equivalent to
$(x_1,x_3,y_2,y_4)=(x_4,x_2,y_3,y_1)$ and in this form it express the
collinearity of the intersection points of the corresponding sides of the
quadrilaterals $X_1X_2Y_2Y_1$ and $X_4X_3Y_3Y_4$. Another equivalent form
$(x_1,x_3,y_4,y_2)=(x_2,x_4,y_3,y_1)$ corresponds to the quadrilaterals
$X_1X_4Y_4Y_1$ and $X_2X_3Y_3Y_2$.

In order to make the symmetry between these three forms more explicit, we
consider the quadrilaterals $X_1X_2X_3X_4$ and $Y_1Y_2Y_3Y_4$ as the
opposite faces of combinatorial cube and rename them as $XX_1X_{12}X_2$ and
$X_3X_{13}X_{123}X_{23}$ respectively. In this notation subscripts
correspond to the coordinate shifts $T_i:X\to X_i$ and their order is
unessential, so that $X_{ij}$ and $X_{ji}$ denote the same point. This
renumeration brings to the double cross-ratio equation \cite{KS1,NS}
\begin{equation}\label{2cr}
  (x,x_{12},x_{13},x_{23})=(x_{123},x_3,x_2,x_1).
\end{equation}
This equation is invariant with respect to any interchange of subscripts and
therefore it express the collinearity of the intersection points of the
corresponding edges for any pair of the opposite faces of the combinatorial
cube.

The double cross-ratio equation is known to be integrable in the sense that
it can be self-consistently embedded into a multi-dimensional lattice
\cite{ABS}. This means the following. Consider (\ref{2cr}) as a partial
difference equation in $\Integer^3$, so that $x=x(n_1,n_2,n_3)$,
$x_1=x(n_1+1,n_2,n_3)$ and so on. Obviously, a generic solution
$x:\Integer^3\to\CP^1$ is uniquely defined by the initial data on the
coordinate planes $n_i=0$. Now, consider the mapping $x:\Integer^M\to\CP^1$,
$M>3$, governed by equation
\begin{equation}\label{xijk}
  (x,x_{ij},x_{ik},x_{jk})=(x_{ijk},x_k,x_j,x_i)
\end{equation}
for any 3-dimensional sublattice. It turns out that such mapping is also
computed from the initial data on the coordinate planes without any
contradictions. In order to prove this, it is sufficient to check that if
the values $x_{123},x_{124},x_{134},x_{234}$ are found from (\ref{xijk})
then equations
\[
  (x_l,x_{ijl},x_{ikl},x_{jkl})=(x_{ijkl},x_{kl},x_{jl},x_{il}),\quad
  \{i,j,k,l\}=\{1,2,3,4\}
\]
define one and the same value $x_{1234}$ as function on the initial data
$x,x_1,\dots,x_{34}$. This property is called {\em 4-dimensional
consistency}. The consistency in the whole $\Integer^M$ follows. In the next
Section we will see that 4-dimensional consistency of double cross-ratio
equation is inherited from some more general construction.

\section{Quadrilateral lattice on the plane}\label{s:pnet}

\begin{figure}[htb]
 \center{\includegraphics[width=120mm]{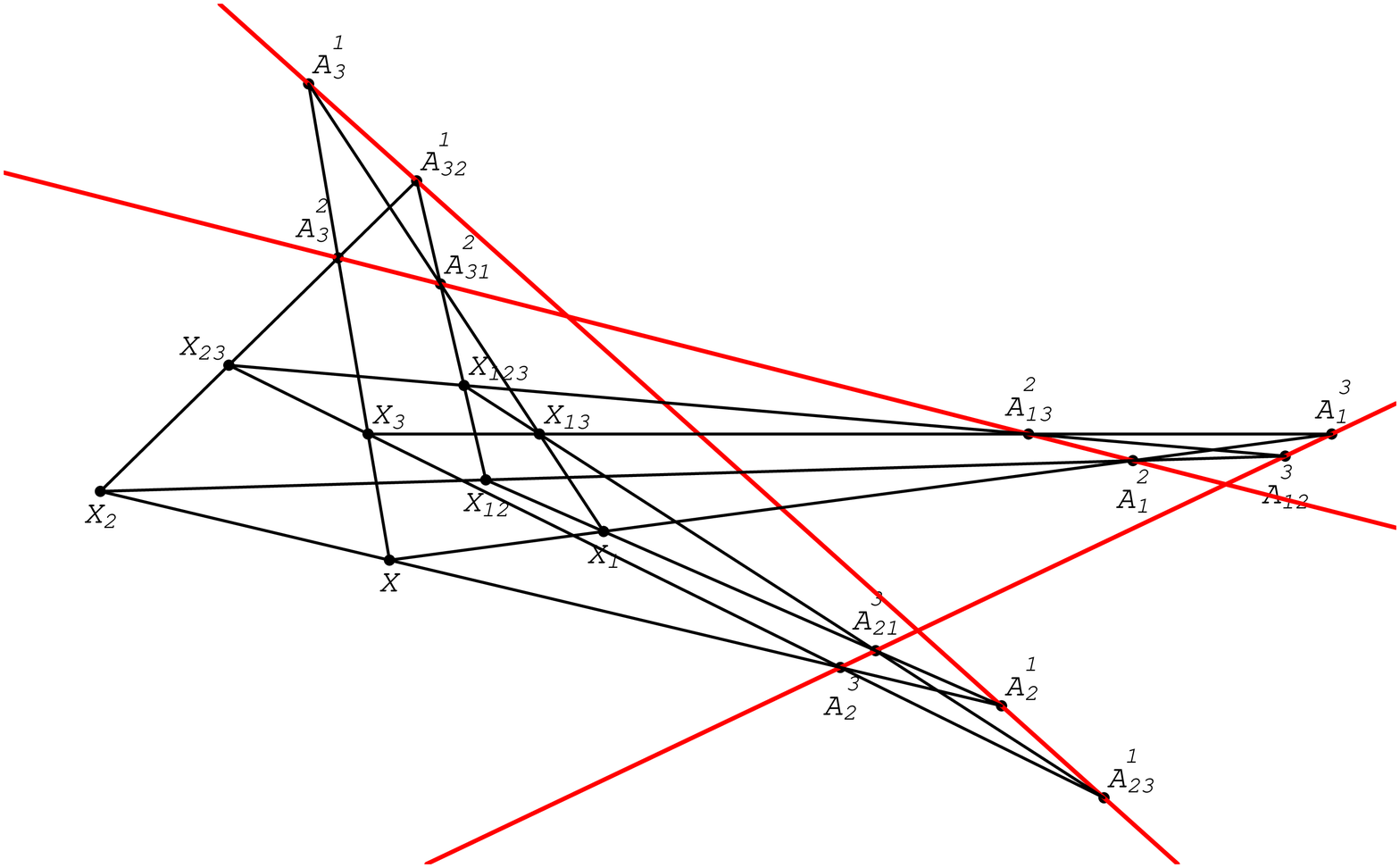}}
 \caption{Theorem \ref{th:60}}\label{fig:60}
\end{figure}

Double cross-ratio equation defines the mapping from 7 into 1 point on the
conic section. If $X,X_1,\dots,X_{23}$ are given then $X_{123}$ is
constructed by means of two most elementary geometric operations only,
drawing a line through two points and finding intersection of two lines:
\begin{equation}\label{71}
\begin{gathered}
   A^3_1=XX_1\cap X_3X_{13},\quad A^3_2=XX_2\cap X_3X_{23} \\
   A^3_{12}=X_2X_{12}\cap A_1A_2,\quad A^3_{21}=X_1X_{12}\cap A_1A_2 \\
   X_{123}= A^3_{12}X_{23} \cap A^3_{21}X_{13}.
\end{gathered}
\end{equation}
This can be considered also as the mapping from 7 into 1 point on the plane.
M\"obius theorem guarantees that if initial data lie on a conic section then
$X_{123}$ lies on it as well. Moreover, in this case $X_{123}$ does not
depend on the interchanging of subscripts in (\ref{71}), due to the symmetry
properties of equation (\ref{2cr}), so that three mappings corresponding to
the different pairs of the opposite faces actually coincide. The natural
question arise, if this is true only for reductions to conic sections, or
also for the generic initial data? The answer is given by the following
theorem (see fig.~\ref{fig:60}).

\begin{theorem}\label{th:60}
Consider a combinatorial cube on the plane. If, for some pair of the
opposite faces, the intersection points of the corresponding edges are
collinear, then the same is true for any other pair.
\end{theorem}
\begin{proof}
As in the case of Desargues theorem, the proof is obtained by constructing a
3-dimensional figure for which the original one is a planar projection.
Denote the intersection points as follows:
\[
  A^k_i = XX_i\cap T_k(XX_i),\quad A^k_{ij}= T_j(XX_i)\cap T_jT_k(XX_i),
  \quad \{i,j,k\}=\{1,2,3\}
\]
and assume that the points $A^3_1,A^3_2,A^3_{12},A^3_{21}$ are collinear.
Let $\alpha$ be the plane of the figure and $\alpha'$ be some other plane
through the line $A^3_1A^3_2$. Choose some point $O$ outside both planes and
define the face $X'_3X'_{13}X'_{123}X'_{23}$ as the projection of the face
$X_3X_{13}X_{123}X_{23}$ from $O$ onto $\alpha'$. All faces of the
combinatorial cube $XX_1X_{12}X_2X'_3X'_{13}X'_{123}X'_{23}$ are planar. For
example, the face $XX_2X'_{23}X'_3$ is planar since the lines $X_3X_{23}$
and $X'_3X'_{23}$ meet in $A^3_2$ by construction and this is also the point
of intersection of the lines $X_3X_{23}$ and $XX_2$.

Now, consider the points
\begin{gather*}
  A^1_2=XX_2\cap X_1X_{12},\quad A'^1_3=XX'_3\cap X_1X'_{13},\\
  A'^1_{23}=X'_3X'_{23}\cap X'_{13}X'_{123},\quad
  A'^1_{32}=X_2X'_{23}\cap X_{12}X'_{123}.
\end{gather*}
They belong to the intersection of the planes $XX_2X'_{23}X'_3$ and
$X_1X_{12}X'_{123}X'_{13}$ and therefore are collinear. Hence, the points
$A^1_2,A^1_3,A^1_{23},A^1_{32}$ which are the projections of these points
from $O$ onto $\alpha$ are also collinear. The collinearity of the points
$A^2_1,A^2_3,A^2_{13},A^2_{31}$ is proved analogously.
\end{proof}

Notice that 8 vertices and 12 sides of the cube, 12 intersection points and
3 lines of intersections form a configuration with the symbol $(20_3 15_4)$.
This configuration is regular, that is, all points and lines are on equal
footing. For example, the lines $A^1_2A^1_{23}$ and $A^2_1A^2_{13}$
correspond to the edges of the combinatorial cube with the opposite faces
$X_2A^1_2X_1A^2_1$ and $X_{23}A^1_{23}X_{13}A^2_{13}$, while the lines
$XX_3$ and $X_{12}X_{123}$ play the role of the lines of intersections for
this cube.

The proof of the Theorem \ref{th:60} makes obvious the link between the
mapping (\ref{71}) and the notion of quadrilateral lattices introduced in
\cite{DS}. Recall that the $M$-dimensional quadrilateral lattice is a
mapping $X:\Integer^M\to\Real^d$, $d>2$, such that the image of any unit
square in $\Integer^M$ is a planar quadrilateral $XX_iX_{ij}X_j$. The image
of any unit cube is a combinatorial cube with planar faces. It is clear that
the vertex $X_{123}$ in such a figure is uniquely defined as the
intersection of the planes
\begin{equation}\label{71'}
 X_{123}=X_1X_{12}X_{13}\cap X_2X_{12}X_{23}\cap X_3X_{13}X_{23}
\end{equation}
and therefore the 3-dimensional quadrilateral lattice is reconstructed from
three 2-dimensional ones corresponding to the coordinate planes which play
the role of initial data. The main property of this mapping proved in
\cite{DS} is its 4-dimensional consistency which guarantees that
$M$-dimensional lattice is also reconstructed from 2-dimensional ones
without contradiction.

Accordingly to the Theorem \ref{th:60}, the projection of quadrilateral
lattice from $\Real^d$ onto a plane is reconstructed from the images of
initial data by applying the mapping (\ref{71}) instead of (\ref{71'}).
Since the quadrilateral lattice in $\Real^d$ is 4-dimensionally consistent,
hence the mapping (\ref{71}) is 4-dimensionally consistent as well. In
particular, this is also true for the reduction of the mapping (\ref{71}) to
a conic section, that is, for double cross-ratio equation (\ref{2cr}).

From all above, the following definition of the quadrilateral lattices on
the plane can be issued.

\paragraph{Definition.} $M$-dimensional quadrilateral lattice on the plane
is a mapping $X:\Integer^M\to\Real^2$, $M\ge3$, such that the images of the
corresponding edges of any pair of the opposite faces of any unit cube in
$\Integer^M$ meet on a straight line.

\section{Hietarinta equation}\label{s:H}

Consider the partial difference equation in $\Integer^2$ introduced by
Hietarinta in the recent paper \cite{H} (up to the change $x\to-x$):
\[
 (x-e^2)(x_1-o^1)(x_2-e^1)(x_{12}-o^2)=(x-e^1)(x_1-e^2)(x_2-o^2)(x_{12}-o^1)
\]
or, as multi-ratio,
\begin{equation}\label{H}
 (x,e^2,x_1,o^1,x_{12},o^2,x_2,e^1)=1.
\end{equation}
Here $e^i,o^i$ are parameters of the equation.

\begin{figure}[htb]
 \center{\includegraphics[width=65mm]{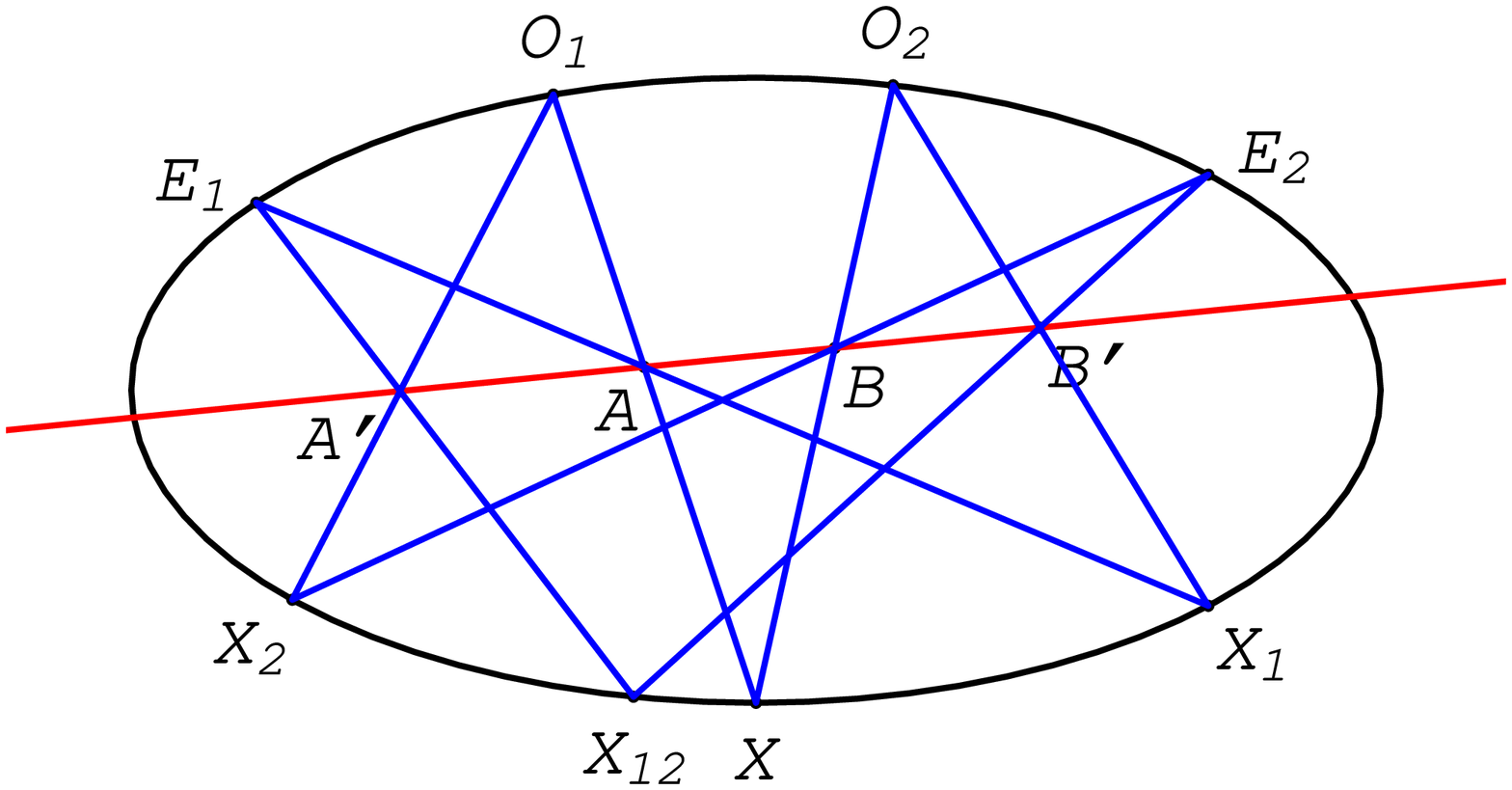}}
 \caption{The mapping (\ref{Hmap})}\label{fig:H}
\end{figure}

Lemma \ref{l:48} provides the geometric interpretation of this equation as
the particular case of Theorem \ref{th:uv} at $N=4$. Like for the double
cross-ratio equation, it is convenient to reformulate this case as a mapping
from 7 into 1 point on a conic section $\CC$. This mapping is illustrated by
the fig.~\ref{fig:H} and is described as follows.

Let $X,X_1,X_2$ and $O^1,O^2,E^1,E^2$ be points on $\CC$. Then the point
$X_{12}\in\CC$ is defined by formulae
\begin{equation}\label{Hmap}
\begin{gathered}
 A=XO^1\cap X_1E^1,\quad B=XO^2\cap X_2E^2 \\
 A'=X_2O^1\cap AB,\quad B'=X_1O^2\cap AB \\
 X_{12}=E^1A'\cap E^2B'.
\end{gathered}
\end{equation}
The corresponding values of the rational parameter on $\CC$ are related by
equation (\ref{H}).

Equation (\ref{H}) is 3-dimensionally consistent, that is, the mapping
$x:\Integer^M\to\CP^1$ governed by equation
\[
 (x,e^j,x_i,o^i,x_{ij},o^j,x_j,e^i)=1
\]
for any 2-dimensional sublattice is computed from the initial data on the
coordinate axes without contradictions. This can be easily checked directly.
At the moment it is not clear, if this property is inherited from some more
general construction, as in the case of double cross-ratio equation, and any
geometric proof is not known.

\paragraph{Acknowledgment.} I thank A.I.~Bobenko and Yu.B.~Suris for
fruitful discussions and many useful suggestions.


\end{document}